\documentclass[10pt,aps,prl,preprintnumbers,twocolumn,nofootinbib]{revtex4-1}
\pdfoutput=1
\usepackage[latin1]{inputenc}
\usepackage{amsmath}
\usepackage{amsfonts}
\usepackage{amssymb}
\usepackage{mathtools}
\usepackage{graphicx}
\usepackage{hyperref}
\usepackage{slashed}
\usepackage[left=1.25cm,right=1.25cm,top=2cm,bottom=2cm]{geometry}
\usepackage{color}

\newcommand{\dd}{\; \mathrm{d}}

\begin{document}

\preprint{CP3-Origins-2017-035 DNRF90}

\title{The Sun as a sub-GeV Dark Matter Accelerator}

\author{Timon Emken}
\email[E-mail: ]{emken@cp3.sdu.dk}
\affiliation{CP${}^3$-Origins, University of Southern Denmark, Campusvej 55, DK-5230 Odense, Denmark}

\author{Chris Kouvaris}
\email[E-mail: ]{kouvaris@cp3.sdu.dk}
\affiliation{CP${}^3$-Origins, University of Southern Denmark, Campusvej 55, DK-5230 Odense, Denmark}

\author{Niklas Gr\o nlund Nielsen}
\email[E-mail: ]{ngnielsen@cp3.sdu.dk}
\affiliation{CP${}^3$-Origins, University of Southern Denmark, Campusvej 55, DK-5230 Odense, Denmark}

\begin{abstract}
\noindent 
Sub-GeV halo dark matter that enters the Sun can potentially scatter off hot solar nuclei and be ejected much faster than its incoming velocity. We derive an expression for the rate and velocity distribution of these reflected particles taking into account the Sun's temperature and opacity. We further demonstrate that future direct detection experiments could use these energetic reflected particles to probe light dark matter in parameter space that cannot be accessed via ordinary halo dark matter.   
\end{abstract}

\maketitle

\section{Introduction} 
Despite convincing astrophysical evidence for the existence of dark matter (DM) in the universe, the numerous direct detection experiments set out to observe DM on Earth have not yet succeeded, severely constraining the classic Weakly Interacting Massive Particle (WIMP) paradigm~\cite{Akerib:2016vxi,Aprile:2017iyp,Cui:2017nnn}. Many recent endeavors target the more elusive sub-GeV DM parameter space~\cite{Angloher:2015ewa,Aguilar-Arevalo:2016ndq,Agnese:2017jvy}, and various ideas have been proposed on how to explore the possibility of light DM~\cite{Essig:2011nj,Graham:2012su,Kouvaris:2016afs}.

Detectors with targets of mass $m_T$ and energy threshold $E_{\rm th}$ can in principle probe DM masses down to $\sim~m_T/\big(\sqrt{2m_Tv_{\rm max}^2/E_{\rm th}}-1\big)$, where $v_{\rm max}$ is the maximum speed of the considered DM population. Below this mass the nuclear recoils caused by even the fastest DM particles are too soft and fall below the experimental threshold, making it impossible to detect DM regardless of exposure and cross section. The two obvious ways to extend sensitivity to lower masses are to use lighter targets and/or realize low energy thresholds as achieved e.g. by CRESST-II~\cite{Angloher:2015ewa}. In the standard halo model one usually assumes $v_{\rm max}=v_\text{gal}+v_{\rm obs}$, the sum of the galactic escape velocity and the velocity of the observer. This value may increase if the standard halo model is extended, e.g. by the inclusion of tidal DM streams from galactic mergers~\cite{Savage:2006qr,Herzog-Arbeitman:2017zbm}. However in the case of sub-GeV DM there is another possibility to generate faster DM particles in the solar system, which has the advantage of being widely halo model independent.

If a light DM particle scatters off a hot nucleus inside the Sun, it will gain energy, and can exit the Sun with a speed far exceeding the incoming one. The velocity after this reflection is no longer limited by the galactic escape velocity. For this to occur at a significant rate, the DM-nucleus interaction must be sufficiently strong, but not so strong, that the outer, colder layers of the Sun shield off the hot and dense core, which accelerates DM particles most efficiently.
Solar reflection becomes effective if the kinetic energy of the infalling DM is smaller than the thermal energy of solar nuclei. 
Hydrogen is the best sub-GeV DM accelerator, because scattering on the lightest solar nuclei results in the largest energy transfers. 
Since the DM velocity deep inside the Sun is dominated by the solar escape velocity, the initial distribution of the incoming halo DM has only a slight impact on the spectrum of the reflected particles. These particles may be much faster than any from the halo and allow high-exposure, low-threshold direct detection experiments to look for lighter DM than naively expected -- potentially setting new constraints on sub-GeV DM, as we will show in this paper.

The idea to search for DM particles accelerated in the Sun was first proposed by one of the authors~\cite{Kouvaris:2015nsa}, who considered DM evaporating to high velocities after being gravitationally captured and thermalized. In contrast to evaporation, reflection does not require thermalization and is thus well defined at an arbitrarily low DM mass.

In this paper, we derive an analytic expression for the scattering and reflection rate. In doing so we generalize the well-known framework by Gould~\cite{Gould:1987ir,Gould:1987ju,Gould:1991hx} beyond the transparent regime. 
In the first section we establish the formalism to describe solar reflection. In the second section we investigate the implications for direct detection. 

Throughout we use natural units, i.e. $\hbar=c=k_\text{B}=1$.

\section{Dark matter scattering in the Sun}

In this section, we derive a formula for the differential rate at which halo DM scatters to final speed $v$ in a spherical shell of the Sun. This formula encapsulates both the reflection and capture rate through single scatterings. The scattering rate of DM in a particular spherical shell consists of three pieces,
\begin{equation}
	\dd S = \dd \Gamma \times \dd P_\text{scat} \times P_\text{shell}\, ,
	\label{Eq: schematic scattering rate}
\end{equation}
where $\dd \Gamma$ is the rate at which free halo DM in an infinitesimal phase space volume would pass through a spherical shell, $\dd P_\text{scat}$ is the probability of scattering off a nucleus in the shell, and finally $P_\text{shell}$ is the probability that the particle reaches the shell without having scattered beforehand. 
We follow in broad strokes the calculations of Press, Spergel and Gould~\cite{Press:1985ug, Gould:1987ir} when evaluating the first two pieces of Eq.~(\ref{Eq: schematic scattering rate}). To describe the Sun's interior structure we use the Standard Solar Model~\cite{Serenelli:2009yc}.

\paragraph{Halo rate into the Sun.}
The rate at which DM reaches the Sun is given by~\cite{Gould:1987ir}
\begin{equation}
\dd \Gamma = \pi n_x f_\text{halo}(u)\frac{\text{d}u\text{d}J^2}{u}\, ,
\label{Eq: rate entering Sun}
\end{equation}
where $n_x$ is the DM number density in the halo, $u$ and $f_\text{halo}$ are the DM speed and speed distribution in the Sun's rest frame  asymptotically far away and  $J$ is the angular momentum per mass of DM with respect to the center of the Sun. For a trajectory to cross the surface of the Sun the angular momentum must be smaller than $J< w(u,R_\odot) R_\odot$; $ R_\odot$ is the solar radius and $w(u,r)$ is the local blueshifted velocity of a DM particle entering a spherical shell of radius $r$, which is $
w(u,r)=\sqrt{u^2+v^2_\text{esc}(r)}$, with $v_\text{esc}(r)$ being the local escape velocity from the Sun. Eq.~(\ref{Eq: rate entering Sun}) should be understood as an average rate, since the Sun's peculiar velocity in the galactic rest frame introduces an anisotropy in the DM flux across its surface.

\paragraph{Scattering rate inside a spherical shell.}
The next piece of Eq.~(\ref{Eq: schematic scattering rate}) is given by 
\begin{equation}
	 \dd P_\text{scat} =  \frac{ \mathrm{ d}l }{w(u,r)} \Omega\left[w(u,r) \right]\, ,
	 \label{Eq: prob of scattering in shell}
\end{equation}
where $\mathrm{ d}l$ is the length traveled in the shell of radial thickness $\mathrm{d}r$, and $\mathrm{d}l/w(u,r)$ is thus the time the particle spends in the shell. Note that $\Omega (w)$ is the total rate of DM scattering off solar nuclei 
\begin{equation}
	\Omega(w) \equiv \sum_N n_N(r)\langle  \sigma_{xN} w_{\text{rel},N}\rangle\, ,
	\label{omega}
\end{equation}
where $\sigma_{xN}$ is the total DM-nucleus scattering cross section, $n_N(r)$ is the number density of nucleus $N$, $ w_{\text{rel},N}$ is the DM-nucleus relative velocity and $\langle \cdot \rangle$ denotes the thermal average. Note that for zero nuclear temperature, $ w_{\text{rel},N} = w$ holds and Eq.~(\ref{Eq: prob of scattering in shell}) reduces to $\text{d}P_\text{scat} = \mathrm{d}l/\lambda(r)$, with $\lambda(r)$ being the DM mean free path\footnote{Previous works \cite{Bernal:2012qh, Garani:2017jcj, Busoni:2017mhe} connected the transparent and opaque regimes as well, where instead of Eq.~\eqref{Eq: prob of scattering in shell} $\text{d}P_\text{scat}$ was taken to be $  \langle \sigma_{xN}\rangle n_N \mathrm{d}l$, and orbits were approximated as straight lines.}. Assuming a velocity independent cross section the relevant thermal average for temperature $T$ is
\begin{equation}
	\langle w_{\text{rel},N}\rangle = \left(1 + 2\kappa^2 w^2 \right)\frac{\text{erf}\left(\kappa w \right)}{2\kappa^2 w} +\frac{e^{-\kappa^2 w^2}}{\sqrt{\pi}\kappa}, \, \kappa\equiv \sqrt{\frac{m_N}{2T}}\, .
\end{equation}\\
We now express $\mathrm{d}l$ in terms of the thickness of the shell $\mathrm{d}r$ by conservation of angular momentum, i.e.
\begin{align}
\frac{\text{d}l}{\text{d}r} = \left[1-\frac{J^2}{w^2(u,r)r^2} \right]^{-1/2}\, .
\end{align}
Finally, we wish to keep information about the spectrum of scattered particles, and therefore we need the differential scattering rate into final velocity $v$ instead of the total scattering rate $\Omega$. This has been computed by Gould under the assumption of isotropic scattering, which is a good approximation, especially in the case of sub-GeV DM~\cite{Gould:1987ju}. It is given by
\begin{widetext}
\begin{equation}
\frac{\dd\Omega}{\dd v}(w \to v)=\frac{2}{\sqrt{\pi}} \frac{v\text{d}v}{w}\sum_N\frac{\mu_+^2}{\mu}\sigma_{xN} n_N(r)\left\{\chi\left(\pm \beta_-,\beta_+ \right)e^{-\mu \kappa^2\left(v^2- w^2 \right)}+ \chi\left(\pm\alpha_-,\alpha_+ \right) \right\}\, ,
\label{Eq: Gould rate}
\end{equation}
\end{widetext}
where the upper sign is for acceleration ($w<v$) and the lower sign is for deceleration ($w>v$). We use Gould's notation,
\begin{align}
\mu&\equiv \frac{m_x}{m_N}\, , \quad \mu_\pm \equiv \frac{\mu \pm 1}{2}\, ,\quad \chi(a,b) \equiv \frac{\sqrt{\pi}}{2}\left( \mathrm{erf} b - \mathrm{erf} a\right)\, ,\notag\\
\alpha_\pm &\equiv\kappa\left(\mu_+ v \pm \mu_- w\right)\, ,\quad \beta_\pm \equiv \kappa\left(\mu_- v \pm \mu_+ w\right)\, ,
\end{align}
with $m_x$ being the DM mass. By integrating over all final velocities, one may explicitly verify that $\Omega(w)$ is  given by Eq.~(\ref{omega}).
\begin{figure*}[t!]
\includegraphics[width=\textwidth]{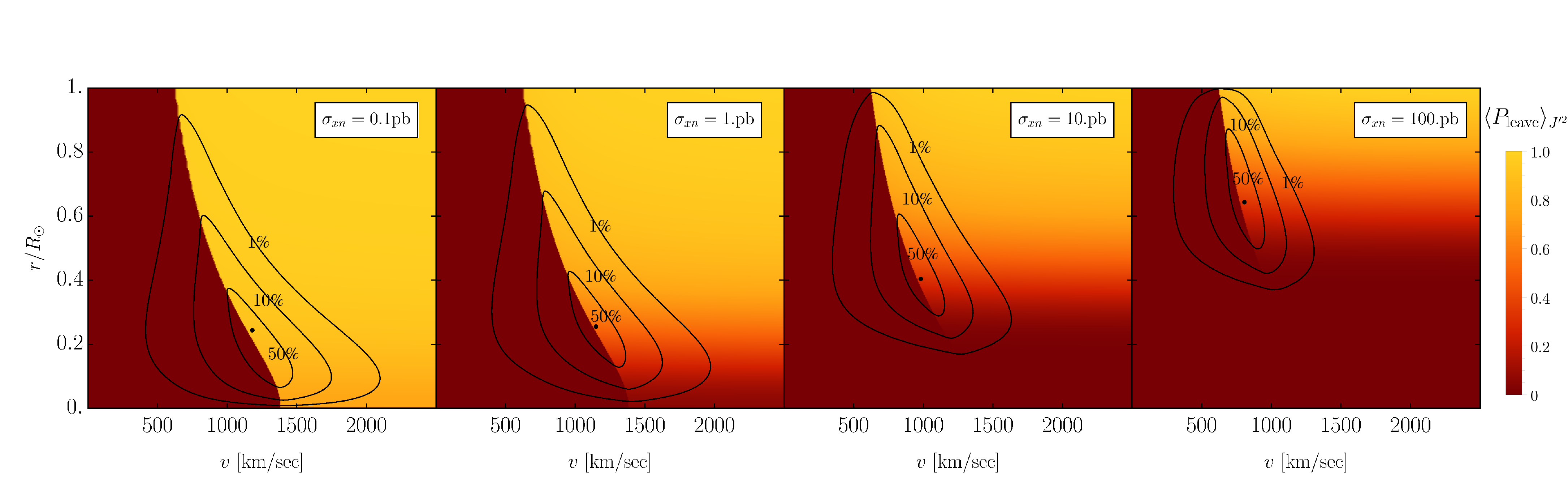}
\caption{The color indicates the probability, $\langle P_\text{leave}(v,r)\rangle_{J'^2}$, for a 200~MeV DM particle with speed $v$ at radius $r$ to exit the Sun without rescattering. Below the escape velocity no particle escapes. 
In addition we superimpose contours of the scattering rate $\text{d}S/\text{d}v\text{d}r$, where the dot denotes the maximum and contours correspond to 50\%, 10\%, and 1\% of the maximal rate. As the Sun becomes more opaque, scatterings peak in the outer colder  layers, which can be seen by the maximum moving to higher $r$ and lower $v$ for larger DM-nucleon cross sections $\sigma_{xn}$.}
\label{Fig: Pleave and dSdvdr}
\end{figure*}
\paragraph{Survival probability of reaching a spherical shell.}
The survival probability when traveling some path in the Sun is $P_\text{surv} = \exp (-\int_\text{path} \mathrm{d}P_\text{scat} )$, where $\mathrm{d}P_\text{scat}$ is given by Eq.~(\ref{Eq: prob of scattering in shell}). For a trajectory, the probability of not scattering between radii $r_A$ and $r_B(>r_A)$ is therefore
\begin{equation}
	P_\text{surv}(r_A,r_B) = \exp\left\{-\int_{r_A}^{r_B}\mathrm{d}r\frac{\mathrm{d}l}{\mathrm{d}r}\frac{ \Omega\left[w(u,r)\right]}{w(u,r)}  \right\}\,.
	\label{Eq: survival probability}
\end{equation}
A free orbit inside the Sun crosses a spherical shell twice or not at all, depending on whether the perihelion is smaller or larger than the radius of the shell. The last term in Eq.~(\ref{Eq: schematic scattering rate}) is thus
\begin{equation}
	P_\text{shell} = P_\text{surv}(r,R_\odot)\left[1+P_\text{surv}^2(r_\text{peri},r)\right]\theta\left[w(u,r)r - J\right]\, ,
	\label{Eq: prob of reaching shell}
\end{equation}
where $r_\text{peri}$ is the perihelion of the orbit and $\theta(x)$ is the step function. The first  and second terms correspond to the survival probabilities of reaching the spherical shell on the first and second passings, respectively.

\paragraph{Final differential scattering rate.}
We are now in a position to combine Eqs.~(\ref{Eq: rate entering Sun}), (\ref{Eq: prob of scattering in shell}) and (\ref{Eq: prob of reaching shell}) with~(\ref{Eq: schematic scattering rate}) to write the differential rate of scattering to final velocity $v$ in a spherical shell of radius $r$,
\begin{widetext}
	\begin{equation}
	\frac{\text{d}S}{\text{d}v\text{d}r} = \pi n_x\int\limits_0^\infty \text{d}u\int\limits_0^{w^2(u,r)r^2} \text{d}J^2  \frac{ f_\text{halo}(u)}{u} P_\text{surv}(r,R_\odot)\left[1+P_\text{surv}^2(r_\text{peri},r)\right]\frac{\text{d}\Omega}{\text{d}v}\left[w(u,r) \to v\right] \left[w^2(u,r)-\frac{J^2}{r^2} \right]^{-1/2}\, .\label{Eq: Scattering rate master equation}
	\end{equation}
\end{widetext}
This expression applies to both the opaque and transparent regimes and smoothly connects the two. 
In general, the integrals must be evaluated numerically. However, it significantly simplifies in the transparent regime, since $P_\text{surv}\approx 1$  and the $J^2$-integral can be evaluated analytically. In this case we obtain an expression similar to the capture rate originally derived by Gould in~\cite{Gould:1987ju}, i.e.
\begin{equation}
		\frac{\text{d}S}{\text{d}v\text{d}r}\approx  4\pi r^2\int_0^\infty \text{d}u  \frac{ f_\text{halo}(u)}{u} w(u,r)\frac{\text{d}\Omega}{\text{d}v}\left[w(u,r) \to v\right] .
\end{equation}
In the completely opaque regime, the formula simplifies again, since $P_\text{surv}(r_\text{peri},R_\odot)\approx 0$ and $P_\text{surv}(r,R_\odot) \times \Omega/w(u,r)\times \mathrm{d}l/\mathrm{d}r \approx \delta(r-R_\odot)$\;\footnote{This can be seen using Eq.~(\ref{Eq: survival probability}), since $P_\text{surv}(r,R_\odot) \times \Omega/w(u,r)\times \mathrm{d}l/\mathrm{d}r$ is a probability density with respect to $r$.}. The total scattering rate can then be evaluated to yield
\begin{equation}
	S\approx \pi R_\odot^2 \left[\langle u\rangle + v_\text{esc}^2(R_\odot)\langle u^{-1}\rangle \right]\, ,
\end{equation}
which by inspection of Eq.~\eqref{Eq: rate entering Sun} is the total rate of DM particles entering the Sun from the halo. In other words, all particles scatter at the surface. In this expression, $\langle u^k \rangle = \int_0^\infty u^k f_\text{halo}(u)\dd u$.

Eq.~\eqref{Eq: Scattering rate master equation} describes the rate at which halo DM scatters once and does not track subsequent scatterings. The effect of multiple scatterings can be described using Monte Carlo simulations, as was recently done in the context of DM-electron scatterings \cite{An:2017ojc}.

\paragraph{Capture and reflection.}
A DM particle that scatters inside the Sun to a new velocity  below the local escape velocity $v_{\rm esc}(r)$ becomes gravitationally bound and is captured.
On the other hand, for a  particle to reflect out  of the Sun after a single scattering, not only does its final velocity need to exceed $v_{\rm esc}(r)$, but it also needs to reach the solar surface without rescattering. We define two probabilities that separate the scattering rate $S$ into a captured and a reflected component, $C$ and $R$, respectively. The probability of capture is simply
\begin{subequations}
	\begin{equation}
P_\text{stay}(v,r) = \theta\left[v_\text{esc}(r)-v\right]\, ,
	\end{equation}
whereas the probability for reflection is
\begin{equation}
	P_\text{leave}(v,r) = \frac{1}{2}  P_\text{surv}(r,R_\odot)\left[1+P_\text{surv}(r'_\text{peri},r)^2\right]\theta\left[v- v_\text{esc}(r)\right]\, .\label{Eq: Pleave}
\end{equation} 
\end{subequations}
Note the similarity to Eq.~(\ref{Eq: prob of reaching shell}). The differential capture and reflection rates are then
\begin{subequations}
\begin{align}
	\frac{\text{d}C}{\text{d}v\text{d}r} &=\frac{\text{d}S}{\text{d}v\text{d}r}P_\text{stay}(v,r)\, ,\\
	\frac{\text{d}R}{\text{d}v\text{d}r} &=\frac{\text{d}S}{\text{d}v\text{d}r}\langle P_\text{leave}(v,r)\rangle_{J'^2}\, ,
\end{align}	
\label{Eq: projection of capture and reflection}
\end{subequations}
where $J'$ is the angular momentum per mass after scattering. As in Eq.~\eqref{Eq: Gould rate}, we assume the scattering to be isotropic when taking the average in the second equation. The behavior of $\text{d}S/\text{d}v\text{d}r$ is shown in Fig.~\ref{Fig: Pleave and dSdvdr} along with $\langle P_\text{leave}(v,r)\rangle_{J'^2}$. We allow scattering on the four largest target nuclei, assuming spin-independent contact interactions: H, $^4$He, $^{16}$O and Fe. Adding further isotopes will slightly increase the overall scattering rate, but also slightly shield the solar core.

Finally, we are interested in the spectrum of reflected particles on  Earth and therefore redshift the reflected particles as
\begin{equation}
	\frac{\mathrm{d} R}{\mathrm{d}u} = \int_0^{R_\odot}\text{d}r \left.\frac{\mathrm{d} R}{\mathrm{d}v\mathrm{d}r}\frac{\mathrm{d}v}{\mathrm{d}u}\right\vert_{v = w(u,r)}\, .
\end{equation}
One might expect the single scattering reflection rate to vanish in the opaque regime, but particles that backscatter near the surface will always have a good chance of making it out again.

\section{Direct detection}
The scattering rate per recoil energy in a detector with $N_\text{T}$ target nuclei has a contribution from the solar reflected DM as well as the standard contribution from the halo,
\begin{equation}
\frac{\text{d}\mathcal{R}}{\text{d}E_\text{R}} = N_\text{T} \int_{u_\text{min}}^\infty \text{d}u\left[\frac{1}{4\pi \ell^2}\frac{\text{d}R}{\text{d}u}+n_x f_\text{halo}(u)u \right] \frac{\text{d}\sigma_{xN}}{\text{d}E_\text{R}}\, .
\end{equation}
Here $\ell =1$~A.U., $n_x = 0.3(\text{GeV}/m_x)$~cm$^{-3}$, $f_\text{halo}$ is the standard isothermal Maxwellian distribution with $v_0=220$~km/sec, $v_\text{obs} = 230$~km/sec and $v_\text{gal}=544$~km/sec. The minimum velocity is given by $u_\text{min} = \sqrt{m_N E_\text{R}/(2\mu_{xN}^2)}$, with $\mu_{xN}$ being the DM-nucleus reduced mass. For spin-independent interactions the differential cross section is
\begin{equation}
\frac{\text{d}\sigma_{xN}^\text{SI}}{\text{d}E_\text{R}} = \frac{m_N A_N^2 \sigma_{xn}^\text{SI}}{2\mu_{xn}^2u^2} F_N^2(E_R)\, ,
\end{equation}
where $n$ refers to nucleons, $A_N$ is the atomic number and $F_N$ is the nuclear form factor. For sub-GeV DM, we can safely set $F_N\approx 1$ for all scatterings in both the Sun and the detector.

\begin{figure}[h!]
\includegraphics[width=0.5\textwidth]{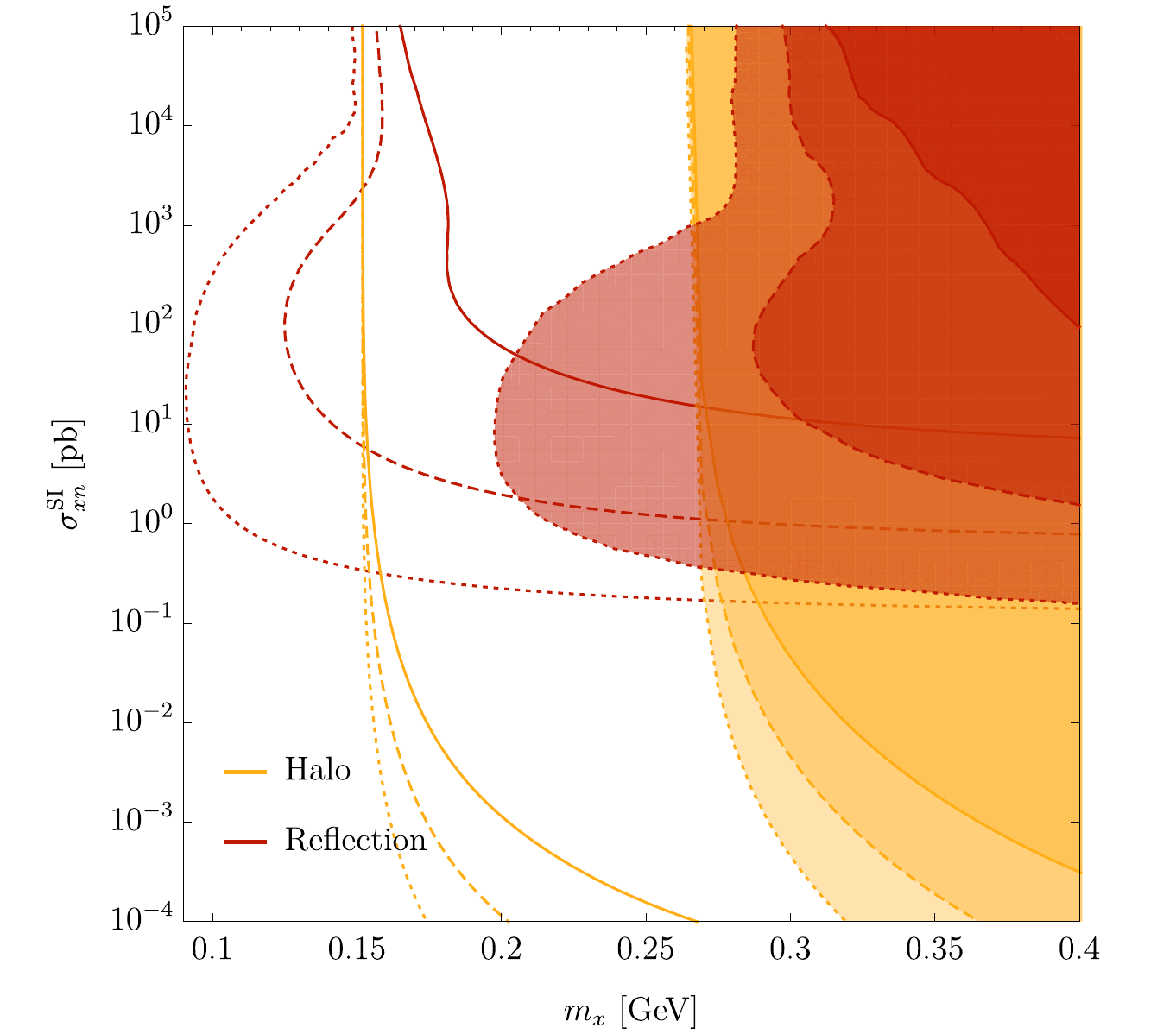}
\caption{The filled contours project constraints for a CRESST-III type detector with exposures of 1/10/100~ton$\cdot$days (solid/dashed/dotted).
The free lines project constraints for an idealized sapphire detector (perfect energy resolution and no background) with 20~eV threshold and exposures of 10/100/1000~kg$\cdot$days (solid/dashed/dotted).
As the exposure increases, halo constraints improve towards lower cross sections only. In contrast, reflection increases the sensitivity to lower masses as well.}
\label{Fig:Constraints}
\end{figure}

The upcoming CRESST-III phase 2 run~\cite{Strauss:2016sxp} will be able to probe solar reflected DM. 
We model the detector response of CRESST-III phase 2 with an expected exposure of 1 ton$\cdot$day according to~\cite{Angloher:2017zkf}. We assume an energy threshold of $E_\text{th}=100$~eV, a Gaussian energy convolution $G( E_\text{R},E_\text{R}')$ with resolution $\sigma_{\text{res}} = E_\text{th}/5$, and efficiencies $\varepsilon_N(E_\text{R}')$ provided by the CRESST Collaboration,
\begin{equation}
\frac{\text{d}\mathcal{R}}{\text{d}E_\text{R}'} =\sum_{N} \varepsilon_N(E_\text{R}')\int_{\tilde{E}_\text{R}}^\infty {\rm d}E_\text{R}\frac{\text{d}\mathcal{R}}{\text{d}E_\text{R}} G( E_\text{R},E_\text{R}')\, ,
\end{equation}
where $N$ runs over O, Ca, and W, and $\tilde{E}_\text{R}=E_\text{th}-2\sigma_\text{res}$.

In Fig.~\ref{Fig:Constraints} we project CRESST-III constraints from solar reflection and halo DM as shaded contours with solid lines at 90\% confidence level~\cite{Yellin:2002xd}. The reflection constraints are subdominant, but with larger exposure solar reflection extends the detector's sensitivity towards lower masses. For a mock detector similar to $\text{CRESST-III}$ with exposures 10 to 100 times larger and no additional background events, a new part of the parameter space becomes accessible (shaded areas with dashed and dotted lines). Using the reflected DM population, a CRESST-III detector with higher exposure can probe  parameter space that is inaccessible with standard halo DM. Furthermore, the reflected component is insensitive to the specifics of the velocity distribution of halo DM, simply because the blueshift of infalling DM and the subsequent scattering erase the memory of the initial distribution.
For experiments with even lower thresholds, reflection can dominate at smaller exposures. For example, in a sapphire detector with a threshold of 20~eV (as demonstrated above ground by the CRESST Collaboration in~\cite{Angloher:2017sxg}), reflection beats halo DM for exposures above $\mathcal{O}(10)$~kg$\cdot$days (see Fig.~\ref{Fig:Constraints}).

{If sub-GeV DM is discovered in the future, solar reflection may be distinguished from (heavier) halo DM in several ways, given sufficient statistics: (1) the recoil spectrum will have a non-Maxwellian tail extending towards high velocities, (2) signals in a directional detector would be pointing towards the Sun, (3) the Earth is about as opaque as the Sun, so one would expect a daily modulation, (4) there would be a $\sim 7\%$ annual modulation due to the eccentricity of the Earth's orbit peak at the perihelion around January 3rd (distinct from halo modulation with a peak around June 2nd).}

\section{Conclusions}
In this paper we demonstrated that above a specific exposure in low threshold direct search experiments, solar reflection of sub-GeV DM can set constraints in the low mass parameter space, where ordinary halo DM cannot. 
The existence of an additional DM population in the solar system with a robust spectrum, insensitive to changes to the halo model, is the central aspect of this work. Our projected constraints are conservative, since our analytic approach accounts only for single scatterings and underestimates the total rate of reflection. Monte Carlo simulations could shed light on the contribution of multiple scatterings. Nonetheless, our formalism is valid in both the opaque and transparent regimes and smoothly connects the two. As a by-product, we improved Gould's expression for the DM capture rate in Eq.~\eqref{Eq: projection of capture and reflection} by taking into account opacity and temperature.

The CRESST-III detector has already achieved thresholds below 100~eV. Solar reflection may therefore be probed by next-generation detectors, pushing the limits of low mass DM searches.

\paragraph{Note:} During the preparation of this manuscript, ref.~\cite{An:2017ojc} appeared, which investigates the prospects of light DM detection via solar reflection as well. It is  complementary to this work, since~\cite{An:2017ojc} focuses on DM-electron interactions using Monte Carlo methods, whereas we focus on DM-nuclei interactions using an entirely analytic approach.

\paragraph{Acknowledgments}
We would like to thank the CRESST Collaboration, and in particular Florian Reindl for helpful comments and assistance regarding CRESST-III. 
This work is partially funded by the Danish National Research Foundation, grant number DNRF90, and by the Danish Council for Independent Research, grant number DFF 4181-00055.

\end{document}